\documentclass[twocolumn,showpacs,aps,pre,a4paper,superscriptaddress]{revtex4-1}
\usepackage[utf8]{inputenc}
\usepackage{amsmath}
\usepackage{amssymb}
\usepackage{siunitx}
\usepackage[english]{babel}
\usepackage{bm}
\usepackage{color}
\usepackage{cleveref}
\crefname{equation}{}{}
\usepackage{etoolbox}
\makeatletter
\appto{\appendix}{%
	\@ifstar{\def\theequation@prefix{A.}}%
	{}%
}
\makeatother

\begin{document}

\title{Relativistic kinetic equation for spin-1/2 particles in the long scale-length approximation} 
\author{R. Ekman}
\affiliation{Department of Physics, Ume{\aa } University, SE--901 87 Ume{\aa}, Sweden}
\author{F. A. Asenjo}
\affiliation{Facultad de Ingeniería y Ciencias, Universidad Adolfo Ibáñez, Santiago, Chile}
\author{J. Zamanian}
\affiliation{Department of Physics, Ume{\aa } University, SE--901 87 Ume{\aa}, Sweden}

\begin{abstract}
In this paper we derive a fully relativistic kinetic theory for spin-1/2 particles and its coupling to Maxwell's equations, valid in the long scale-length limit, where the fields vary on a scale much longer than the localization of the particles; we work to first order in $\hbar$.
Our starting point is a Foldy-Wouthuysen (FW) transformation, applicable to 
this regime, of the Dirac Hamiltonian.
We derive the corresponding evolution equation for the Wigner quasi-distribution in an external electromagnetic field.
Using a Lagrangian method we find expressions for  the charge and current densities, expressed as free and bound parts.
It is furthermore found that the velocity is non-trivially related to the  momentum variable, with the difference depending on the spin and the external  electromagnetic fields.
This fact that has previously been discussed as ``hidden momentum'' and is due to that the FW transformation maps pointlike particles to particle clouds for which the prescription of minimal coupling is incorrect, as they have multipole moments.
We express energy and momentum conservation for the system of particles and the electromagnetic field, and discuss our results in the context of the Abraham-Minkowski dilemma.
 
\end{abstract} 

\maketitle

\section{Introduction}

In both high intensity laser-matter experiments and in astrophysical settings, the spin of the electron may play a significant role~\cite{Hu1999,Walser2000,Walser2002,Mahajan2014}.
It is thus necessary to find a relativistic description of spin-$\frac 1 2$ electrons in strong electromagnetic fields.
While such a description is provided by the Dirac equation~\cite{Dirac1928a}, it is fully consistent only in the context of quantum field theory~\cite{*[{See for example the discussion in }] [{, Ch.~1.}] WeinbergI} where pair production is incorporated.
However, if pair production is negligible, the Foldy-Wouthuysen transformation~\cite{foldy1950dirac,Silenko2008,Chen2014a} separates particles and anti-particles in an expansion in $\hbar$, and a particle theory can be constructed.

In this paper, we derive a scalar fully relativistic kinetic equation for spin-$\frac 1 2$ particles in electromagnetic fields, starting from the Foldy-Wouthuysen transformation.
We emphasize that the theory presented here is fully relativistic, that is, applicable to all orders in $v/c$, in contrast to previous $O(v^2/c^2)$ semi-relativistic work~\cite{asenjo2012semi}.
By considering a mean-field Lagrangian, we obtain expressions for the charge and current densities, as the sum of free and bound parts.
We show that coupled to Maxwell's equations, the system fulfills an energy conservation law and identify the Poynting vector.

The Foldy-Wouthuysen transformation carries with it some difficulties of interpretation concerning observables~\cite{Chen2014}, tracing from the problem of localized particles in relativistic theories~\cite{RevModPhys.21.400,PhysRev.87.688}, and the ``hidden momentum'' of systems with magnetic moments~\cite{*[{See }]  [{ for a review, and references therein.}] Babsonetal2009}.
In particular, while it is clear that there must be a contribution to the current density from the spin and its form in the Dirac theory is well known ~\cite{gurtler1975consistency,VanHolten1991}, the corresponding expression in the Foldy-Wouthuysen representation has previously been found only in the semi-relativistic ($O(v^2/c^2)$) limit~\cite{asenjo2012semi,PhysRevA.88.032117}.
Based on these points, we briefly discuss our result in the context of the Abraham-Minkowski dilemma.

An alternative to the Foldy-Wouthuysen transformation is the Frenkel model~\cite{Frenkel1926,Frenkel1926a,Ternov1980}, which models the spin as a classical angular momentum.
The Frenkel and Foldy-Wouthuysen models have recently~\cite{Wen2016} been numerically benchmarked against each other and the Dirac equation for single-particle motion in strong laser pulses, showing disagreement in experimentally accessible regimes.
However, more work needs to be done to investigate the validity of the classical model, and especially concerning collective effects.
To this end, a kinetic theory based on the Frenkel model will be discussed in a forthcoming paper.

\section{The Long Scale-Length Hamiltonian} 

\subsection{Foldy-Wouthuysen transformation and domain of applicability}

As the starting point we will take the relativistic Hamiltonian derived by Silenko~\cite{Silenko2008} using a Foldy-Wouthuysen transformation.
A necessary condition for the Foldy-Wouthuysen transformation to be valid is that the process of pair creation is negligible.
We will give a more precise condition below.

The Dirac Hamiltonian is of the form 
\begin{equation}
	\mathcal H = \beta m + \mathcal E + \mathcal O 
\end{equation}
where $m$ is the mass, $\mathcal E = q\phi$, and $\mathcal O = \boldsymbol{\alpha}\cdot (\mathbf p - q\mathbf A)$.
Here $q$ is the charge, $\phi, \mathbf A$ the scalar and vector potentials, $\mathbf p$ the momentum operator, and $\beta, \boldsymbol{\alpha}$ are the Dirac matrices.
In this form, $\mathcal E$ ($\mathcal O$) is called the \emph{even} (\emph{odd}) parts of the Hamiltonian, as it commutes (anti-commutes) with $\beta$: $[\beta, \mathcal E] = 0$ ($\{ \beta, \mathcal O\} = 0$).
It is the odd part that connects the upper two components of the bispinor with the lower two, thus coupling positive and negative energy states and allowing for pair creation.

For the free Dirac Hamiltonian, Foldy and Wouthuysen~\cite{foldy1950dirac} found the transformation that renders the odd part exactly zero.
In external fields, the transformation and thus the transformed Hamiltonian can be found as an expansion in powers of the potentials and their derivatives.
In this expansion, the parameter that should be small in order to truncate the expansion is the particle localization $l$ compared to the inhomogeneity size $L$ of the external potentials, that is $ L \gg l$ \cite{Silenko2003,Silenko2013}.
Since $\hat{\mathbf x}$ in the Foldy-Wouthuysen representation is the Newton-Wigner~\cite{RevModPhys.21.400} position operator, $l \sim \hbar/E$ where $E^2 = m^2 + p^2$.
Thus $l$ is the Compton wavelength for low-energy particles, and the de Broglie wavelength for high-energy particles.
In particular, the fields cannot be comparable to the critical field $E_\text{crit} = \SI{1.3e18}{V/m}$.

\newcommand{\hateps}{\ensuremath{\hat{\epsilon}}}

To first order in the fields, the transformed Hamiltonian for the upper components is~\cite{Silenko2008}
\begin{widetext}
\begin{align}
	\hat H = & 
	\hateps + q \phi 
	- \frac{\mu_B m}{2}  
	\left\{ 
	\frac{ 1}{\hateps} 
		, 
		\bm \sigma \cdot \mathbf B 
	\right\}_+
	+ \frac{\mu_B m }{ \sqrt{2 \hateps (\hateps + m)} }
	\left[ \bm \sigma \cdot (\hat{\bm \pi} \times \mathbf E - \mathbf E \times \hat{\bm \pi}) \right] 
	\frac{1}{\sqrt{2 \hateps(\hateps+ m )} } \label{eq:Hamiltonian}
\end{align}
\end{widetext} 
where we have defined $\hateps \equiv \epsilon(\hat{\bm \pi}) = \sqrt{ m^2 + \hat{\bm \pi}^2 }$ with the kinetic momentum operator given by $\hat{\bm \pi} = \hat{\mathbf p} - q \mathbf A (\hat{\mathbf x})$ (to be discussed in more detail below), and $\bm \sigma$ is a vector containing the Pauli matrices as components.
We use units where $c = 1$ and furthermore have denoted the particle mass by $m$, charge by $q$ and we have the Dirac magnetic moment $\mu_B = q \hbar / 2m $.
The external fields are given by $\mathbf E = - \nabla \phi - \partial_t \mathbf A$ and $\mathbf B = \nabla \times \mathbf A$, where $\phi$ and $\mathbf A$ are respectively the scalar and vector potentials, and $\left\{ A , B \right\}_+ = AB + B A$ is the anticommutator.
The third and fourth term in the Hamiltonian above represent the relativistic interaction of the spin with the electromagnetic field.

The Hamiltonian above is derived to lowest order in $\hbar$, but, again, is fully relativistic, provided the condition $L \gg l$ is fulfilled.
In Ref.~\cite{Silenko2008} a more general Hamiltonian for a particle with anomalous magnetic moment and an electric dipole moment was derived, but we will here only consider a particle with the Dirac magnetic moment and we hence have set the $g$-factor to 2.
Kinetic effects of the anomalous magnetic moment were investigated, non-relativistically, in Ref.~\cite{brodin1}.  

\subsection{Observables and their corresponding operators}
There are subtleties in connecting operators in the Foldy-Wouthuysen representation to physical observables \cite{gurtler1975consistency, VanHolten1991, Chen2014}.
For example, the velocity operator of the particle is given by the time derivative of the position operator in the Heisenberg picture.
Using the Heisenberg equation of motion we get 
\begin{widetext}
\begin{equation}
	\dot{\hat{\mathbf x}} =  \, \frac{i}{\hbar} \left[ \hat H , \hat{\mathbf x} \right] 
	= \frac{1}{2} \left\{ \frac{1}{ \hat{\epsilon}} , \hat{\bm \pi} \right\}_+ 
 	+ \frac{\mu_B m}{\hat{\epsilon}^3} (\bm \sigma \cdot \mathbf B) \hat{\bm \pi}
	- \frac{ \mu_B m ( 2 \hat{\epsilon} + m) }{\hat{\epsilon}^3 ( \hat{\epsilon} + m)^2 } 
	\left[ \bm \sigma \cdot (\hat{\bm \pi} \times \mathbf E - \mathbf E \times \hat{\bm \pi}) \right] \hat{\bm \pi} - \mu_B m \frac{\mathbf E \times \bm \sigma - \bm \sigma \times \mathbf E}{2\hat{\epsilon}(\hat{\epsilon}+m)} , 
	 \label{eq:velocityOperator} 
\end{equation}
\end{widetext} 
to lowest order in $\hbar$.
The first term is the usual relation between the velocity and the momentum, while the rest of the terms are due to the spin and the external field.
In particular, the momentum variable $\hat{\bm \pi}$ is no longer co-linear with the velocity vector.
This is already the case for the spin-orbit interaction to order $O(v^2/c^2)$, where the last term in~\eqref{eq:velocityOperator} also appears~\cite{*[{}] [{, Sec.~33.}] Berestetskii1982}.
The two other correction terms are of the same order in the mass but contain a factor $\hat{\bm \pi}$; they are therefore relativistic corrections.

The evolution equation of the spin is
\begin{widetext}
\begin{equation}
	\dot{\bm \sigma} = \bm\sigma\times \frac{\mu_B}{m\hbar}\left[ \left\{ \frac{1}{\hateps}, \mathbf B\right\}_+ - \frac{1}{\sqrt{\hateps(\hateps+m)}}\big[\bm \pi \times \mathbf E - \mathbf E \times \bm \pi \big]\frac{1}{\sqrt{\hateps(\hateps+m)}}\right]
\end{equation}
\end{widetext}
whence we can easily see that the length of the polarization $\bm \sigma$ is a constant.
Furthermore, the classical limit of this is precisely the classical equation of motion for the rest frame spin~\cite{Jackson}. 
It is thus clear that the operator $\bm \sigma$ in the Foldy-Wouthuysen representation gives the polarization of the electron in the electron rest frame~\cite{Chen2014,fradkin1961,Silenko2003}.

Based on this it is possible to understand the form of the Hamiltonian~\eqref{eq:Hamiltonian}.
It is of course well known that the canonical momentum and the mechanical momentum are not equal,
but also the prescription of \emph{minimial coupling}, $\bm \pi = \mathbf{p} - q\mathbf A$ is not universally valid.
Indeed if this minimal coupling prescription is the physically correct relation in one representation it will not remain valid after a momentum-dependent canonical transformation -- such as the Foldy-Wouthuysen transformation \cite{Costella:1995gt}.

The origin of the $\bm\sigma\times\mathbf E$ contribution to the kinetic momentum, as in~\eqref{eq:velocityOperator}, can be understood by realizing, as Foldy and Wouthuysen did~\cite{foldy1950dirac} that the $\hat{\mathbf x}$ operator in the Foldy-Wouthuysen representation is a mean position operator~\cite{RevModPhys.21.400}.
Its eigenstates are smeared out combinations of eigenstates of the Pauli-Dirac position operator.
The latter states are individually point electric charges, but their superpositions have dipole (and higher) moments.
The Foldy-Wouthuysen particle is thus a current loop magnetic dipole, possessing ``hidden momentum''~\cite{Babsonetal2009,PhysRevLett.20.343,PhysRev.128.2454}, which if unaccounted for leads to apparent violations of Newton's third law in the Shockley-James paradox~\cite{PhysRevLett.18.876,PhysRev.171.1370}.

\section{The Gauge Invariant Wigner and Spin Transformations} 
The density operator $\hat \rho$ for the system evolves according to the von Neumann equation which is given by 
\begin{equation}
	i \hbar \dot{\hat{\rho}} = \left[ \hat H , \hat \rho \right] \label{eq:vonNeumann}.  
	\end{equation} 
This evolution equation can be transformed into an evolution equation for the Wigner quasi-distribution function~\cite{wigner32}.
This form resembles the classical Vlasov equation and methods from this theory can straightforwardly be applied to the quantum version.
However, an important difference is that the Wigner function is not necessarily non-negative -- hence not a proper classical probability distribution -- since it has to respect the Heisenberg uncertainty principle. 

\subsection{The Wigner-Stratonovich transformation} 
Since we are dealing with electromagnetic fields, we need to use the gauge invariant version of the Wigner function first derived by Stratonovich~\cite{stratonovich1956gauge}.

The Wigner-Stratonovich transformation is obtained by 
\begin{align}
	W_{\alpha \beta} ( \mathbf x, \mathbf p, t) = &  \int \frac{d^3 \mathbf z}{(2\pi \hbar)^3} 
	e^{ \frac{ i}{\hbar} \mathbf z \cdot \left[ \mathbf p + q \int_{-1/2}^{1/2} d\lambda 
	\mathbf A(\mathbf x + \lambda \mathbf z , t) \right] }
\notag \\ & 
	\times \rho_{\alpha \beta} \left( \mathbf x + \frac{\mathbf z}{2} , \mathbf x - \frac{\mathbf z}{2} , t \right) 
	\label{wigner-stratonovich1956gauge}
\end{align}
where the integral in the exponent is the Wilson loop that makes $W_{\alpha\beta}$ gauge invariant.
To calculate the evolution equation, we write out the von Neumann equation~\eqref{eq:vonNeumann} in terms of $\rho(\mathbf x, \mathbf y)$,
and functions of the operators $\bm{\pi}_x = -i\hbar \nabla_x - q \mathbf A(\mathbf x), \bm{\pi}_y = -i\hbar \nabla_y - q \mathbf A(\mathbf y)$ acting on $\rho(\mathbf x, \mathbf y)$.
We use the identities~\cite{stratonovich1956gauge}
	\begin{subequations}
\begin{align}
	\begin{array}{c}
	{\mathcal F} \left[ \bm{\pi}_x \right] \rho(\mathbf x, \mathbf y)
	 \mapsto
	{\mathcal F} \bigg[ \mathbf p - \frac{i\hbar}{2} 
	\nabla_x \\ + \frac{i\hbar q }{2} \int_{-1}^1 d\eta \frac{1 + \eta}{2} 
	\mathbf B \left( \mathbf x + \frac{\eta i\hbar}{2} \nabla_p \right ) \times \nabla_p
	\big] W(\mathbf x, \mathbf p)
	\end{array}
	\label{eq:identityI} \\
	\begin{array}{c}
	\mathcal F \left[{\bm \pi}_y \right] \rho(\mathbf x, \mathbf y)
	\mapsto
	\mathcal F \bigg[ \mathbf p + \frac{i\hbar}{2} 
	\nabla_y - \\
	\frac{i\hbar q }{2} \int_{-1}^1 d\eta \frac{1 - \eta}{2} 
	\mathbf B \left( \mathbf y + \frac{\eta i\hbar}{2} \nabla_p \right ) \times \nabla_p
	\bigg] W(\mathbf x, \mathbf p)
	\end{array}
	\label{eq:identityII} 
\end{align} 
\end{subequations}
where we have approximated the expressions to first order in $\hbar$ assuming again that the scale-length of the fields and potentials varies very little over the extensions of the particle wave functions.
The last expansion is obtained by using the formula~\cite{kumar} 
\begin{equation}
	f(A + B) \approx f(A) + \frac{1}{2} \left\{ f' (A) , B \right\} , 
\end{equation}
where $A$ and $B$ are operators (in general noncommuting) and the expansion is to first order in $B$ which is considered small compared to $A$.  

The first identity~\eqref{eq:identityI} is derived in Ref.~\cite{stratonovich1956gauge}, and the others are derived in a similar fashion from the definition~\eqref{wigner-stratonovich1956gauge}. 

\subsection{The spin transform} 
The quasidistribution function $W_{\alpha\beta}$ is matrix-valued, where the matrix components correspond to the different spin states and their transition probabilities.
To get a scalar distribution function we apply the spin transform~\cite{ZamanianNJP} given by 
\begin{align}
	f(\mathbf x, \mathbf p , \mathbf s , t) &=  
	\frac{1}{4\pi} \textrm{Tr} 
	\left[ ( 1 + \mathbf s \cdot \bm \sigma) W(\mathbf x, \mathbf p) \right] 
	\notag \\ &
	= \frac{1}{4\pi}  \left( \delta_{\alpha \beta} + \mathbf s \cdot \bm \sigma_{\alpha \beta} 
	\right) W_{\beta \alpha} ( \mathbf x, \mathbf p , t ) , 
\end{align}
where a summation over $\alpha, \beta = 1,2$ is understood in the last equality and $\mathbf s$ is a unit vector. In terms of this new distribution function, the expectation value of the polarization operator is given by  
\begin{equation}
	\left< \bm \sigma \right>(t) = 3 \int d^3 \mathbf x d^3 \mathbf p d^2 \mathbf s f(\mathbf x, \mathbf p, \mathbf s, t) \mathbf s ,
\end{equation} 
where the factor 3 is there to compensate for the fact that the quantum mechanical distribution is always spread out in $\mathbf s$, even for a completely polarized particle~\cite{ZamanianNJP}.

For the spin transformation it is straightforward to deduce the identities
\begin{subequations}
\begin{align} 
 \int \nabla_s  f \, d^2 \Omega  & = 2 \int \mathbf s f \, d^2 \Omega \label{eq:spinIdentityI}  \\
 \int \mathbf s \times \mathbf X \cdot \nabla_s f  \, d^2 \Omega& = 0 \label{eq:spinIdentityII}
\end{align} 
\end{subequations}
where $\mathbf X$ is any vector independent of $\mathbf s$.
These identities are important in deriving the results in Section~\ref{sec:properties}.

\subsection{The kinetic equation} 
Starting from the evolution equation for the density matrix~\eqref{eq:vonNeumann}, performing the Wigner-Stratonovich transformation 
using the identities~\eqref{eq:identityI}--\eqref{eq:identityII}, 
and then taking the spin transformation, and using the identities~\eqref{eq:spinIdentityI}--\eqref{eq:spinIdentityII} and we get 
\begin{widetext} 
\begin{align}
	0  & = \partial_t f + 
	\left\{ 
		\frac{\mathbf p}{\epsilon} - \mu_B m \nabla_p 
			\left[ 
			\frac{1}{\epsilon} 
				\left( 
				\mathbf B - \frac{ \mathbf p \times \mathbf E}{\epsilon + m} 
				\right) 
				\cdot 
				\left( 
				\mathbf s + \nabla_s 
				\right) 
			\right] 
	\right\} 
	\cdot \nabla_x f 
	\notag \\ &
	+ q \left(
		\mathbf E + 
			\left\{ 
			\frac{ \mathbf p}{\epsilon} - \mu_B m 	\nabla_p 
				\left[ 
				\frac{1}{\epsilon} 
					\left( 
					\mathbf B - \frac{ \mathbf p \times \mathbf E}{\epsilon + m}
					\right) 
					\cdot \
					\left( 
					\mathbf s + \nabla_s
					\right)
				\right]
			\right\} 
			\times \mathbf B
		\right)
		\cdot \nabla_p f
	+ \frac{\mu_B m}{\epsilon} \nabla_x 
	\left[ 
		\left( 
		\mathbf B - \frac{\mathbf p \times \mathbf E}{\epsilon + m} 
		\right) 
		\cdot 
		\left( 
		\mathbf s + \nabla_s 
		\right) 
	\right] 
	\cdot \nabla_p f
	\notag \\ &
	+ \frac{2 \mu_B m }{\hbar \epsilon} 
	\left[ 
	\mathbf s \times 
		\left( 
		\mathbf B - \frac{\mathbf p \times \mathbf E}{\epsilon + m} 
		\right) 
	\right] \cdot \nabla_s f
	\label{eq:evolution} 
\end{align} 
\end{widetext} 
This is our main result for this paper: the relativistic kinetic equation for spin-1/2 particles in the long-scale length limit. 
Compared to the non-relativistic evolution equation in Ref.~\cite{ZamanianNJP}, \eqref{eq:evolution} modifies the relation between momentum and velocity, and adds to the magnetic dipole interaction ($\sim \mathbf B \cdot \mathbf s$) the spin-orbit interaction ($\sim (\mathbf p\times \mathbf E)\cdot\mathbf s$).

\section{Coupling to the electromagnetic field -- mean-field theory} \label{meanfield} 

In the mean-field approximation, the fields are assumed to be in the form $\mathbf E = \mathbf{E}_{\textrm{ext}} + \mathbf{E}_{\textrm{s.c.}}$ where the first part is an external field and the second part is a self-consistent mean-field.
The mean-fields are assumed to be given by solving Maxwell's equations 
\begin{subequations}
\begin{align}
	\nabla\cdot\mathbf E & = \rho_f - \nabla\cdot\mathbf P \\ 
	\nabla \cdot \mathbf B & = 0 \\
	\nabla \times \mathbf E & = - \frac{\partial \mathbf B}{\partial t} \\
	\nabla\times \mathbf B & = \mathbf j_f + \frac{\partial \mathbf E}{\partial t} + \frac{\partial \mathbf P}{\partial t} + \nabla\times \mathbf M
\end{align}
\end{subequations}
where $\mathbf P$ and $\mathbf M$ are the polarization and magnetization, and $\rho_f$ and $\mathbf j_f$ are the free charge and current densities.

\subsection{Finding the sources from the Lagrangian}

It is not entirely obvious what should be the taken as the sources of the electromagnetic field in Maxwell's equations. One previous paper~\cite{PhysRevA.88.032117} treated the semi-relativistic (second order in $p/m$) case by using a Lagrangian formalism where the sources are simply read off from the Euler-Lagrange equations for the electromagnetic field. The same results appear in other works~\cite{HussainStefanBrodin,stefan2013linear,wang2006consistency}.

A difficulty that appears when using a Lagrangian formalism is the operator  $\epsilon = (m^2 + (\hat{\mathbf p} - iq\mathbf A)^2)^{1/2}$, which contains  derivatives of $\mathbf A$ to all orders, since $\hat{\mathbf p}$ is a differential operator 
We can, however, circumvent this in the mean-field  approximation by passing to the Wigner-Weyl formulation of quantum mechanics,  where $\mathbf p$ is just half of the coordinates in phase-space.
This is in fact natural since we are interested in a kinetic theory: the sources will automatically be expressed in terms of moments of the Wigner function $f$.

We propose that to the action for the free electromagnetic field, we should add $-\int dt \langle H \rangle $, where $\langle H \rangle $ is the expectation value of the Hamiltonian operator~\eqref{eq:Hamiltonian}, \begin{equation}
\langle H \rangle =  \operatorname{Tr}[\hat H \hat \rho] = \int d^3p d^3x d^2s \, H f
\end{equation}
where $H$ is the function on phase space corresponding to $\hat H$.
The Euler-Lagrange equations will then give Maxwell's equations with sources.
We comment further on this procedure at the end of this section.

The function $H(\mathbf x, \mathbf p, \mathbf s)$ is found by writing the Hamiltonian~\eqref{eq:Hamiltonian} such 
that the operators $\hat {\mathbf x}$ and $\hat{\mathbf p} - iq\hat{\mathbf A}$ 
appear in totally symmetric order~\cite{serimaa}, using the commutation 
relations if necessary (note that the fields are functions of $\hat{\mathbf 
	x}$), then putting $\hat{\mathbf{x}} \mapsto \mathbf x$ and $\hat{\mathbf{p}} - 
iq\hat{\mathbf A} \mapsto \mathbf p$.

This reordering is only necessary for the terms in~\eqref{eq:Hamiltonian} that 
are already proportional to $\hbar$.
Reordering operators using commutation relation in these would give terms 
proportional to $\hbar$ and containing a derivative on the fields.
Finding $H$, then, to the order we are interested in, is as simple as removing 
hats from~\eqref{eq:Hamiltonian}.

With these considerations, the mean-field Hamiltonian is explicitly
\begin{equation}
\langle H \rangle = \int d^3x\, d\Omega \,
	\left[
		\epsilon + q \phi
		- \frac{\mu_B m}{\epsilon} 3 \mathbf s \cdot
		\left(\mathbf B  - \frac{\mathbf p \times \mathbf E}{ (\epsilon + m) } \right)
	\right] f \label{eq:MFHamiltonian}
\end{equation}
where $\epsilon^2 = \mathbf p^2 + m^2$ and we have introduced $d\Omega = d^3 p d^2s$ for brevity.

Since the Euler-Lagrange equations are linear in the action, we find Gauss's 
law as \begin{equation}
\nabla\cdot\mathbf E = q \int d\Omega \, f + \nabla\cdot \int d\Omega 
\frac{3\mu_B m}{\epsilon(\epsilon+m)} \mathbf s \times \mathbf p \,  f. \label{eq:Gauss}
\end{equation}
This has the interpretation of a free charge density $\rho_f$ minus the divergence of a polarization field, $\nabla\cdot\mathbf P$.
In the non-relativistic limit, we reproduce the previous  result of Refs.~\cite{PhysRevA.88.032117} and~\cite{asenjo2012semi}.

For Amp\`{e}re's law, we have to be careful and remember that $f$ is  gauge-invariant only because of the Wilson line factor entering into the  definition~\eqref{wigner-stratonovich1956gauge}.
In computing the variation with respect to the vector potential, the variation  of the Wilson line factor must also be taken into account. Viz.,
\begin{equation}
	\frac{\delta}{\delta A_i} \int d\Omega\,  O f
	= \int d\Omega\, \bigg(
		\frac{\delta O}{\delta A_i} f + O \frac{\delta f}{\delta A_i}
	\bigg) .
	\label{eq:variation}
\end{equation}
From the definition of the gauge-invariant Wigner function, one finds 
\begin{equation}
	\frac{\delta f}{\delta A_i} = q\delta(x-x')\nabla_p f
\end{equation}
so that after an integration by parts, the variation is
\begin{equation}
	\frac{\delta}{\delta A_i} \int d\Omega O f = \int d\Omega \left (\frac{\delta 
	O}{\delta A_i} f - q(\nabla_p O) f \right) \delta(x-x').
\end{equation}

Performing this variation, we find Amp\`{e}re's law as
\begin{widetext}
	\begin{equation}
	0 = \frac{\delta S}{\delta A_i} = 
	-\nabla\times \mathbf B + \frac{\partial \mathbf  E}{\partial t}
	+ q\int d\Omega \, \left[\frac{\mathbf p}{\epsilon} - \mu_B m\nabla_p \frac{1}{\epsilon}\bigg(\mathbf B - \frac{\mathbf p \times \mathbf  E}{\epsilon + m}\bigg)\cdot 3\mathbf s\right]f
	+  \nabla\times \int d\Omega \, \frac{\mu_B  m}{\epsilon}3\mathbf s f
	- \partial_t \int d\Omega \frac{3\mu_B m}{ \epsilon(\epsilon+m)} \mathbf s \times\mathbf p f. \label{eq:Ampere}
	\end{equation}
\end{widetext}

We can interpret the source terms as, in order, $\mathbf j_f$, the free current density; $\nabla\times\mathbf M$, the curl of the magnetization density; and $\partial_t \mathbf P$, the polarization current density.
The semi-relativistic limit again agrees with previous results~\cite{PhysRevA.88.032117,asenjo2012semi}.
For reference, we state explicitly that \begin{align}
\mathbf P  & = -3\mu_B  \int d\Omega \, \frac{m\mathbf s \times \mathbf p}{\epsilon(\epsilon+m)} \label{eq:polarization} \\
\mathbf M & = 3\mu_B \int d\Omega\, \frac{m}{\epsilon} \mathbf s f. \label{eq:magnetization}
\end{align}

The bound charge and current satisfy a conservation law for purely algebraic reasons: \begin{equation}
\partial_t \nabla\cdot\mathbf P = \nabla\cdot(\partial_t \mathbf P + \nabla\times \mathbf M)
\end{equation}
since $\operatorname{div}\operatorname{curl} = 0$. 
We will demonstrate below, using only the evolution equation for $f$, that the continuity equation $\partial_t \rho_f + \nabla\cdot\mathbf j_f = 0$ holds, so that \cref{eq:evolution,eq:Gauss,eq:Ampere} form a consistent system.
(The remaining two of Maxwell's equations are, of course, identically true in the potential formulation.)

\subsection{Some comments}

One might wonder why one cannot simply add $\langle H \rangle$ to the  \emph{Hamiltonian} of the free electromagnetic field.
The reason for not doing so is that while for the free electromagnetic field,  the momentum conjugate to $A_i$ is $E_i$, we cannot expect this to hold in the  presence of the spin-orbit interaction which is proportional to $\mathbf s  \cdot (\mathbf p \times \mathbf E)$.
Thus naively using Hamiltonian methods, evaluating Poisson brackets by means of  \begin{equation}
\{ A_i(x), E_j(x')\} \overset{!}{=} \delta_{ij} \delta(x-x')  \label{eq:naiveHamilton}
\end{equation} will produce inconsistent results.
(For example, if we had  used~\eqref{eq:naiveHamilton} we would not have found the polarization in~\eqref{eq:Gauss}.)
There is more structure to Hamiltonian mechanics than just the Hamiltonian function.
Having found the momentum conjugate to $A_i$ in the presence of the spin-orbit  coupling, however, one can reproduce equivalent equations of motion in 
Hamiltonian form, provided one is sufficiently careful about electrodynamics  being a \emph{constrained} Hamiltonian system, for which we should apply the methods of Dirac~\cite{DiracLectures,10.2307/100496}.

While there is admittedly some arbitrariness in the procedure just described, it is borne out by that it defines a consistent system that, as we will show below, has an energy conservation law, and that the semi-relativistic limits are all correct.


\section{Properties of the model} \label{sec:properties}
\subsection{Continuity equation and the velocity operator}

It can be seen that the number density of particles $n$ is given by $ n= \int d\Omega\, f $.  Taking the zeroth moment of the evolution equation~\eqref{eq:evolution}, remembering to use the spin integral identity~\eqref{eq:spinIdentityI}, we find that it can be written in the form
\begin{equation}
	0 = \partial_t n + \nabla_x\cdot \int  d\Omega\, \mathbf v f \label{eq:continuity}
\end{equation}
where \begin{equation}
	\mathbf v = \frac{\mathbf p}{\epsilon} - \mu_B m\nabla_p 
				\bigg(\frac{\mathbf B}{\epsilon} - \frac{ \mathbf p \times \mathbf E}{\epsilon(\epsilon + m)}\bigg)\cdot 3\mathbf s. \label{eq:velocityFunction}
\end{equation}
The function $\mathbf v$ on phase space is in fact precisely the function that is in Weyl correspondence with $\dot{\hat{\mathbf{x}}}$ as given by~\eqref{eq:velocityOperator}, remembering that $\hat{\bm\sigma} \mapsto 3\mathbf s$ when using the scalar $f$.
Since the Wigner-Weyl phase space formulation of quantum mechanics is equivalent to the Hilbert space formulation
, it should not be surprising that the non-trivial relation between velocity and momentum appears also in this formulation.

Knowing that this function corresponds to the velocity, the kinetic equation~\eqref{eq:evolution} becomes somewhat easier to interpret.
It is in fact analogous to the Vlasov equation.
The same discussion and interpretation as above in connection with~\eqref{eq:velocityOperator} applies.
Furthermore,~\eqref{eq:continuity} multiplied by $q$ is precisely the continuity equation for the free charge and current, in~\eqref{eq:Gauss} and~\eqref{eq:Ampere}.

\subsection{Conservation of energy and the Abraham-Minkowski dilemma}

The system of Maxwell's equations with polarization and magnetization given by \cref{eq:polarization,eq:magnetization}, and the kinetic equation~\eqref{eq:evolution} can be shown to have an energy conservation law, of the form \begin{equation}
\partial_t W + \nabla\cdot \mathbf K = 0 \label{eq:conservation}
\end{equation}
where the total energy density is given by \begin{equation}
W = \frac{1}{2}(E^2 + B^2) + \int d \Omega \, \left (\epsilon - 3\mu_B m \frac{\mathbf B}{\epsilon} \cdot \mathbf s\right)f
\end{equation}
and the energy flux vector is
\begin{multline}
\mathbf K =  
\int \left[ \epsilon + \mu_B m 3\mathbf s \cdot \left(\frac{\mathbf B}{\epsilon} - \frac{\mathbf p\times \mathbf E}{\epsilon(\epsilon+m)}\right) \mathbf v \right] f \,  d\Omega  
\\+  \mathbf E \times \mathbf H 
\label{eq:momentumVector}
\end{multline}
where $\mathbf H = \mathbf B - \mathbf M$.
Again, the semi-relativistic limit agrees with previous results~\cite{asenjo2012semi}.


In~\eqref{eq:momentumVector}, the Poynting vector appears to be $\mathbf E \times \mathbf H$.
Some comments on this, in the context of the Abraham-Minkowski dilemma, are in order. 
As is well known (see, e.g., the review in Ref.~\cite{Griffiths2012}) there are two proposals for the electromagnetic momentum density vector, the Minkowski momentum density \begin{equation}
\mathbf g_M = \mathbf D \times \mathbf B
\end{equation}
where $\mathbf D = \mathbf E + \mathbf P$, and the Abraham momentum density \begin{equation}
\mathbf g_A = \mathbf E \times \mathbf H
\end{equation}
where, again, $\mathbf H = \mathbf B - \mathbf M$.
In a relativistic theory with symmetric stress-energy tensor, if the the energy conservation law is of the form~\eqref{eq:conservation}, then $\mathbf K$ must be the momentum density and should be conserved, \begin{equation}
\partial_t K_i + \partial_i T_{ij} = 0
\end{equation}
for some symmetric tensor $T_{ij}$ ($i,j = 1,2,3$).
Our Eqs.~\eqref{eq:conservation}--\eqref{eq:momentumVector} would then suggest that the Abraham alternative is correct.
However, there is more to be said.

In non-relativistic work~\cite{GransSamuelsson2015} using the kinetic theory from Ref.~\cite{ZamanianNJP}, it was found that \begin{multline}
\partial_t (\mathbf D \times \mathbf B + \mathbf p) + \partial_i T_{ij} = 0
\\ = \partial_t (\mathbf E \times \mathbf H + \mathbf p) + \partial_i T_{ij} + \partial_t (\mathbf E \times \mathbf M)
\end{multline}
where $\mathbf p$ is the particle momentum density.
This would seem to favor Minkowski.
However, since that work was non-relativistic -- lowest order in $v/c$ -- the Hamiltonian, and thus evolution equation, does not include the $O(v^2/c^2)$ spin-orbit interaction.
Consequently, the relation between particle momentum and velocity is the standard $\mathbf p = m \mathbf v$, that is, the hidden momentum is neglected, and the hidden momentum is, to $O(v^2/c^2)$, precisely $\mathbf E \times \mathbf M$.
This agrees with the argument by Barnett~\cite{PhysRevLett.104.070401} that the dilemma is between canonical momentum (defined by $[\hat x, \hat p_\text{can}] = i\hbar$) and kinetic momentum, $\gamma m \mathbf v$.

However, working to lowest order in $v/c$, the Shockley-James paradox~\cite{PhysRevLett.18.876} appears.
In this paradox, there is a force from an uncharged magnet with changing magnetic moment on a test charge, due to the induced electric field, but no obvious force on the magnet, as required by Newton's third law.
Coleman and van Vleck's resolution~\cite{PhysRev.171.1370} of the Shockley-James paradox by arguing that a theory including electromagnetic momentum must be at least $O(v^2/c^2)$ to be consistent.
The argument is that if a non-static $\mathbf B$ is generated by the movement of charge carriers, it is $O(v/c)$.
The induced electric field, determined by Faraday's law, is also $O(v/c)$.
Since the electromagnetic momentum is quadratic in the fields, all terms of order $v^2/c^2$ must then be kept.
In the Shockley-James case, this means taking the gamma factor $\gamma = 1 + v^2/2$ in $\mathbf p = m\gamma\mathbf v$ as the magnetic moment is entirely due to free current.

Filho and Saldanha~\cite{Filho2015} recently gave a similar, but quantum mechanical argument, demonstrating the presence of hidden momentum in various states of the hydrogen atom.
They, however, choose to not include the spin, and so could not ``deduce if [spin] has or does not have hidden momentum''.
The forms of observables in the Foldy-Wouthuysen representation is strong support for hidden momentum associated with spin.

We hope to in future work investigate the semi-relativistic limit in detail, to confirm that Barnett's~\cite{PhysRevLett.104.070401} argument mends the problems of Ref.~\cite{GransSamuelsson2015}.

\section{Discussion} 


In this paper we have presented a kinetic theory that is both fully relativistic and includes the electron spin.
Previously, at most one of these has been included~\cite{Akama1970,ZamanianNJP}; thus the present model is of theoretical interest, for completeness sake.
In addition, we believe it clarifies the Abraham-Minkowski dilemma, highlighting the connection with hidden momentum.

As for practical applications, strong magnetic fields are very important in the realm of relativistic quantum mechanics.
In plasmas, if a magnetic field is strong enough, such that the gyroradius of constituents is of the order of the de Broglie length, then quantum effects become relevant.
In this case, the spin of the particles affects the plasma dynamics.
Thus, a semi-classical kinetic theory, as the one constructed here, is useful to obtain the first-order quantum corrections to descriptions of strongly magnetized plasmas.
Examples are pulsar atmospheres~\cite{Asseo2003,Gurevich1985,Harding2006} or high-energy laser-plasmas~\cite{Tatarakis2002}, where quantum effects have been found to be very relevant at high energies~\cite{Melrose2002,Melrose2011,Gonoskov2015,Harvey2009,DiPiazza2012}.

Since the theory presented here is based on separating positive and negative energy states of the Dirac equation, it cannot describe pair production.
This is a limit on the field strengths allowed: it means that $\xi_0 = |e|E/m\omega \ll 1$ where $\omega$ is the frequency of the fields.
For optical systems, $\omega \approx \SI{1}{eV}$, this strictly means intensities in the $\SI{e17}{W/m^2}$ range, but the process is strongly kinematically suppressed even with intensities an order of magnitude higher, and electron energies in the tens of GeVs~\cite[Sec.~VIII]{DiPiazza2012}.
We have also discarded derivatives of the fields, which is justified only if the fields vary little over the particle localization distance.
This places a limit at least as strict as $L \gg \lambda_C = \hbar/(mc)$ with $\lambda_C$ the Compton wavelength and $L$ the lengths-scale of the fields.

One further limitation of the theory presented here is that it does not include effects of radiation reaction (RR), which has vexed electrodynamics for over a century (see, e.g., Refs.~\cite{DiPiazza2012,Burton2014} for recent reviews).
The contribution of a magnetic moment to RR has been studied classically~\cite{Bhabha1940,Bhabha1941}.
Very recently, a paper particularly relevant to us~\cite{Wen2016a} compared the Frenkel and Foldy-Wouthuysen models (for one-particle motion) including radiation reaction for both using the Landau-Lifshitz equation.
In principle, the Landau-Lifshitz force could be added by hand to~\eqref{eq:evolution} as a $\mathbf{F}_\text{LL}\cdot \nabla_p f$ term, but studying this extension is beyond the scope of the present paper.

Radiation reaction, however, is quantitatively small except in the most extreme regimes, with the relative importance of the RR and Lorentz forces being $\eta = \alpha \gamma^2 E/E_\text{crit}$ where $\gamma$ is the Lorentz factor and $\alpha$ the fine-structure constant.
Intensities high enough that $\eta \approx 1$ are only expected to be reached with next-generation laser facilities~\cite{DiPiazza2012}, and by inspection
there are clearly regimes where $\eta \ll 1$, but $\gamma$ is large enough that an $O(v^2/c^2)$ treatment is inapplicable.

In summary, while the model we have presented here discards some physics, we believe it may still be useful for systems such as those mentioned at the beginning of this section.

\begin{acknowledgments} 
	The authors are grateful for helpful email correspondence with A. J.~Silenko, and to two anonymous referees for providing comments that helped improve the paper.
	R.~E. was supported by by the Swedish Research Council, grant number 2012-3320.
	J.~Z. acknowledges financial support from the Wallenberg Foundation within the grant ``Plasma based compact ion sources'' (PLIONA).
\end{acknowledgments}

\appendix*

\section{Equations of motion in Hamiltonian formalism}

\newcommand{\bA}{\mathbf{A}}
\newcommand{\bE}{\mathbf{E}}
\newcommand{\bB}{\mathbf{B}}
\newcommand{\bp}{\mathbf{p}}
\newcommand{\bs}{\mathbf{s}}
In this appendix, we show explicitly that the same Maxwell equations are found using a constrained Hamiltonian systems approach, as described by Dirac~\cite{DiracLectures}.

The field variables in the Lagrangian are the scalar potential $\phi$ and the vector potential $\bA$, with corresponding momenta \begin{subequations}
	\begin{align}
		\pi_\phi & = 0 \\
		\pi_\bA = \frac{\partial L}{\partial \dot{\bA}} = - \frac{\partial L}{\partial \bE}
			& = -\bE -\int d\Omega \, \frac{3\mu_B \bp \times \bs}{\epsilon(\epsilon+m)} \, f \label{eq:pi-E}.
	\end{align}
\end{subequations}
Note that $\pi_\bA = -\bE - \mathbf{P} = -\mathbf{D}$.
Just as for the particles, the canonical momentum is modified to contain a spin-dependent term.
For brevity, we let $\mu' = \frac{3\mu_B m}{\epsilon(\epsilon + m)}$.
Since $\dot \phi$ does not appear in the Lagrangian, the Lagrangian is singular and the first of these is a constraint.
Now, the naive (i.e., without Lagrange multipliers) Hamiltonian \emph{density} is
\begin{widetext}
\begin{align}
	\mathcal H_0 & = \pi_\bA \cdot \dot{\bA} - L
				  = \big (\bE + \int d\Omega \, \mu' \bp \times \bs \, f\big) \cdot (\bE + \nabla \phi) - \frac{E^2 - B^2}{2}
				  + H_\text{MF}
				  \notag \\
				  & = \frac{E^2 + B^2}{2}  + \nabla \phi \cdot
					 \bigg(\bE +  \int d\Omega \, \mu' \bp \times \bs\, f \bigg)   
					 + \int d\Omega \left[ \epsilon + q \phi - \frac{3\mu_B}{\epsilon}\bs \cdot \bB \right] f
\end{align}
\end{widetext}
and the corresponding Hamiltonian is $H_0 = \int d^3 x \, \mathcal H_0$.
Note that since the Hamiltonian is a function of the fields and the momenta, $\bE$ should be expressed according to \eqref{eq:pi-E}, i.e., it \emph{it is independent of $\phi$}. (This is because treating electrodynamics with Hamiltonian methods, the scalar potential corresponds to gauge freedom, as will be seen below.)

Now, we check if there are any secondary constraints and find Gauss's law:
\begin{widetext} \begin{align}
 0 & \approx \dot{\pi_\phi} = \{\pi_\phi, H_0 \} = \frac{\delta H_0}{\delta \phi(x)} = 
	 \int d^3 x' \, \left [ (\nabla \delta)(x - x') \cdot \bigg(\bE +   \int d\Omega \, \mu' \bp \times \bs  f  \bigg)
	 + \int d\Omega \, q \delta(x - x') f
	 \right ]  \notag\\
	& = - \nabla \cdot \bE + q\int d\Omega \, f -
		\nabla \cdot \int d\Omega \, \frac{3\mu_B m}{\epsilon (\epsilon + m)} \bp \times \bs \,  f.
\end{align}\end{widetext}
Clearly, the free and bound charge densities are the same as we found with the Euler-Lagrange equations, namely~\eqref{eq:Gauss} and~\eqref{eq:polarization}.
Since $\phi$ does not appear in the secondary constraint, the constraints are first class, i.e., they correspond to gauge freedom.

To see if there are any tertiary constraints, we must consider the time evolution of $\bE$, which will give us Amp\`{e}re's law.
By using~\eqref{eq:pi-E}, and identifying the $\bp \times \bs$ term as the polarization, the time evolution of $\bE$ is given by \begin{equation}
	\begin{aligned}-\partial_t \bE  = &  
		\{ p_\bA, H_0 \}		
		+ \partial_t \mathbf{P}, \\
		&  + \big\{\int d\Omega \, \mu' \bp \times \bs\,  f, H_0 \big\}
	\end{aligned}
\end{equation}
For the first Poisson bracket, using~\eqref{eq:variation} we find \begin{widetext}
\begin{align}
 \{ p_\bA , H_0 \}  & = \frac{\delta H_0}{\delta \bA(x)} =  -\nabla \times \bB 
		  - q\int d\Omega \, \nabla_p (  \nabla \phi \cdot \mu' (\bp \times \bs) f )  \notag \\
		 & + \int d^3 x' \int d\Omega \, \left[  \frac{3 \mu_B}{\epsilon} \bs \times (\nabla \delta)(x- x')  - q \nabla_p (\epsilon + \frac{3 \mu_B}{\epsilon} \bs \cdot \bB ) \delta(x-x')\right] f \notag\\
	 & = - \nabla \times \bB - q\int d\Omega \, \bigg[ \frac{\bp}{\epsilon} + 3\mu_B  \nabla_p \bigg ( 
		 \frac{\bB}{\epsilon} + \frac{\nabla \phi\times \bp}{\epsilon(\epsilon+m)}
	\bigg) \cdot \bs
	\bigg] f + \nabla \times \mathbf M.
\end{align}
\end{widetext}
where the magnetization $\mathbf M = \int d\Omega \, \frac{3\mu_B\bs}{\epsilon} f$ just as we found before in~\eqref{eq:magnetization}.
For the second Poisson bracket we again use~\eqref{eq:variation} and obtain (summation over $i$ implied)
\begin{widetext}
 \begin{align}
	\big\{\int d\Omega \, \mu' \bp \times \bs\,  f, H_0 \big\}  & = 
		\int d^3 x' \, d^3 x'' \, \left( \frac{\delta}{\delta A_i(x'')}
		\int d\Omega \, \mu' \bp \times \bs f
		 \right )
		 \frac{\delta \mathcal H_0(x')}{\delta {\pi_A}_i (x'')} \notag \\
		 &= \int d^3x'  d^3x'' d\Omega \, \big( (\nabla_{p_i} \mu' \bp \times \bs) \,  \delta(x-x'') f \big) (E_i + \nabla_i \phi ) \delta(x-x') \notag  \\
		  &= \int d\Omega \,  \nabla_p \bigg( \frac{3 \mu_B \bp \times \bs}{\epsilon(\epsilon + m)}\cdot (\bE + \nabla \phi) \bigg) f .
\end{align}
\end{widetext}

We collect terms and conclude that
\begin{widetext}
	\begin{equation}
		\partial_t \bE = \nabla\times \bB - q \int d\Omega \left[ \frac{\bp}{\epsilon} - 3\mu_B  \nabla_p \bigg ( 
		\frac{\bB}{\epsilon} - \frac{\bp \times \bE}{\epsilon(\epsilon+m)}
		\bigg) \cdot \bs\right] f - \nabla\times \mathbf M  - \partial_t \mathbf{P}
	\end{equation}
\end{widetext}
with the same free current, magnetization $\mathbf M$, and polarization $\mathbf P$ as using the Lagrangian approach.

In principle, we should follow Dirac and add Lagrange multipliers that enforce the constraint $p_\phi = 0$ to the Hamiltonian density.
However, because the constraint would be relevant only for the equation of motion for $\phi$, which merely generates gauge transformations,
we need not be concerned with the time evolution of $\phi$, and the above analysis gives the correct Maxwell equations.

Checking for tertiary constraints we get \begin{equation}\begin{aligned}
 \partial_t (\nabla \cdot \bE)  \approx \nabla\cdot(\nabla\times (\bB - \mathbf M) - \partial_t \mathbf P - \mathbf j_\text{f}) \\
   \approx \partial_t \rho_\text{f} - \partial_t \nabla\cdot \mathbf P,\end{aligned}
 \end{equation}
i.e., the continuity equation for the bound charge, which is~\eqref{eq:continuity}, found using only the evolution equation for the Wigner function $f$.
Therefore, there are no tertiary constraints.


\bibliography{../../Bibliographies/Relativistic_kinetic_equations_with_spin}{}

\end{document}